\documentstyle[aps,epsf]{revtex}
\begin{document}
\pagestyle{empty}
\begin{flushright}
FERMILAB-PUB-99/218-E\\
\today
\end{flushright}
\footnotetext[1]{Submitted to Phys. Rev. D}


\vspace{1.0cm}
\begin{center}
{\Large\bf Search for antiproton decay at
the Fermilab Antiproton Accumulator
\footnotemark[1]}
\end{center}
\vspace{0.5cm}

\begin{center}
\renewcommand{\baselinestretch}{1}
S. Geer, J. Marriner, M. Martens, R.E. Ray, J. Streets, W. Wester \\
{\it Fermi National Accelerator Laboratory, Batavia, Illinois 60510}\\
\vspace{0.2cm}
M. Hu, G.R. Snow \\
{\it University of Nebraska, Lincoln, Nebraska 68588}\\
\vspace{0.2cm}
T. Armstrong \\
{\it Pennsylvania State University, University Park, Pennsylvania 16802}\\
\vspace{0.2cm}
C. Buchanan, B. Corbin, M. Lindgren, T. Muller$^{\dag}$ \\
{\it University of California at Los Angeles, Los Angeles, California 90024} \\
\vspace{0.2cm}
R. Gustafson \\
{\it University of Michigan, Ann Arbor, Michigan 48109}\\
\end{center}

\vspace{0.1in}
\begin{center}
(The APEX Collaboration)
\end{center}
\begin{abstract}
\nopagebreak
A search for antiproton decay 
has been made at the Fermilab Antiproton Accumulator. Limits are 
placed on thirteen antiproton decay modes. The results include 
the first explicit experimental 
limits on the muonic decay modes of the antiproton, and the first 
limits on the decay modes $e^-\gamma\gamma$ and $e^-\omega$.
The most stringent limit is for the decay mode 
$\overline{p} \rightarrow e^-\gamma$. At 90\% C.L. we find that 
$\tau_{\,\overline{p}}$/$B(\overline{p}\rightarrow e^-\gamma)
> 7 \times 10^5$~yr. 
The most stringent limit for decay modes with a muon in the 
final state is for the decay $\overline{p} \rightarrow \mu^-\gamma$. 
At 90\% C.L. we find that
$\tau_{\,\overline{p}}$/$B(\overline{p}\rightarrow \mu^-\gamma)
> 5 \times 10^4$~yr.
\end{abstract}


\clearpage
\pagestyle{plain}
\setcounter{page}{1}
%
%
\section{Introduction \protect\label{intro}}

The CPT theorem requires that the proton and antiproton ($\overline{p}$)
lifetimes are equal. Searches for proton decay have yielded lower limits on the
proton lifetime $\tau_p > O(10^{32})$~yr~\cite{protondecay}. Hence, a search
for $\overline{p}$ decay with a short lifetime ($\tau_{\,\overline{p}} <<
\tau_p$) tests the CPT theorem. In this paper we describe a search for
$\overline{p}$ decay at the Fermilab Antiproton Accumulator.

CPT invariance is one of the most fundamental symmetries of modern physics.
Over the years there have been a variety of
searches~\cite{protondecay,cptlimits,ion,k0limits} for CPT violation based on
comparing particle and antiparticle masses, lifetimes, and magnetic moments.
For example, the proton and antiproton masses have been shown to be equal 
with a precision of a few parts in $10^{8}$~\cite{ion}, whilst the particle
and antiparticle masses in the neutral kaon system have been shown to be equal
to about one part in $10^{18}$~\cite{protondecay,k0limits}. A search for
$\overline{p}$ decay complements these CPT tests by providing a search for CPT
violation accompanied by a violation of baryon number. Indeed, since the
antiproton is the only long--lived antiparticle that could in principle decay
into other known particles without violating charge conservation, a search for
$\overline{p}$ decay provides a unique test of the CPT theorem, and a unique
test of the intrinsic stability of antimatter. 

The sensitivity of a $\overline{p}$ decay search to the presence of a CPT
violating interaction has been characterized~\cite{crprl} by considering a
dimension-6 CPT-violating operator with characteristic mass scale $m_X$.
Dimensional analysis then provides the estimate $\tau_{\,\overline{p}} \sim
m_X^4/m_p^5$, yielding: 
\begin{equation}
m_X \sim (4.3 \times 10^9 \; GeV) \times (\tau_{\,\overline{p}}/10\;Myr)^{1/4}.
\protect\label{mxlimit}
\end{equation}
Hence, a search for $\overline{p}$ decay with a lifetime of order 1~yr, for
example, provides a test for CPT violation at a mass scale O($10^7$)~GeV/c$^2$,
which is well beyond the scale accessible at high energy colliders.

The most stringent lower limit on $\tau_{\,\overline{p}}$ has been
obtained~\cite{crprl} from a comparison of recent measurements of the cosmic
ray $\overline{p}$ flux~\cite{crays} with predictions based on expectations for
secondary production of antiprotons in the interstellar medium. The agreement
between the observed and predicted rates implies that $\tau_{\,\overline{p}}$
is not small compared to T/$\gamma$, where T is the $\overline{p}$ confinement
time within the galaxy ($\sim 10^7$~yr) and $\gamma$ is the Lorentz factor for
the observed antiprotons. After taking into account the uncertainties on the
relationship between the interstellar $\overline{p}$ flux and the flux observed
at the Earth, at 90\% C.L. the limit $\tau_{\,\overline{p}} > 8 \times 10^5$~yr
has been reported~\cite{crprl}. This indirect limit is not valid if current
models of $\overline{p}$ production, propagation, and interaction in the
interstellar medium are seriously flawed. Indeed, it has been
claimed~\cite{susy} that, within the framework of the minimal supersymmetric
extension to the standard model, if neutralino annihilations contribute
significantly to the flux of cosmic ray antiprotons in the galactic halo,
$\overline{p}$ lifetimes as low as $10^5$~years can accommodate the current
cosmic ray data.

Laboratory searches for $\overline{p}$ decay have, to date, provided less
stringent limits on $\tau_{\,\overline{p}}$. However, these limits do not
suffer from large model dependent uncertainties. The most stringent published
laboratory limit on inclusive $\overline{p}$ decay has been obtained from a
measurement of the containment lifetime of $\sim 1000$ antiprotons stored in an
ion trap, yielding $\tau_{\,\overline{p}} > 3.4$~months~\cite{ion}. The
sensitivity of laboratory $\overline{p}$ decay searches can be dramatically
improved by looking for explicit $\overline{p}$ decay modes at an antiproton
storage ring. Angular momentum conservation requires that a decaying
$\overline{p}$ would produce a fermion (electron, muon, or neutrino) in the
final state. Hence, there are only a limited number of potential decay modes. 
A search for explicit $\overline{p}$ decay modes with an electron in the final
state was made by the APEX test experiment (T861) at the Fermilab Antiproton
Accumulator. The T861 search yielded the 95\% C.L. limits~\cite{t861}:
$\tau_{\,\overline{p}}/B(\overline{p} \rightarrow e^-\gamma) > 1848$~yr,
$\tau_{\,\overline{p}}/B(\overline{p} \rightarrow e^-\pi^0) > 554$~yr,
$\tau_{\,\overline{p}}/B(\overline{p} \rightarrow e^-\eta) > 171$~yr,
$\tau_{\,\overline{p}}/B(\overline{p} \rightarrow e^-K^0_S) > 29$~yr, and
$\tau_{\,\overline{p}}/B(\overline{p} \rightarrow e^-K^0_L) > 9$~yr.

Following the T861 test experiment, the APEX detector~\cite{proposal,nim} was
designed to enable a more sensitive search for $\overline{p}$ decay at the
Fermilab Antiproton Accumulator. Based on APEX data, we have recently reported
limits on $\tau_{\,\overline{p}}/B$ from the first search for antiproton decay
modes with a muon in the final state~\cite{martin,muonprl}. In this paper we
give a more detailed description of our limits on $\overline{p}$ decay modes
with a muon in the final state, and describe new results from a search for
$\overline{p}$ decay modes with an electron in the final state. 

The paper is organized as follows. Sec.~II gives a brief
description of the APEX detector. The APEX triggers and data samples are
described in Sec.~III. The search for decay modes with a muon in the 
final state is described in Sec.~IV., and the electron decay mode
search is presented in Sec.~V. Conclusions are summarized in
Sec.~VI.

\section{The APEX detector \protect\label{detector}}

\begin{figure}
\epsfxsize6.in
\centerline{\epsffile{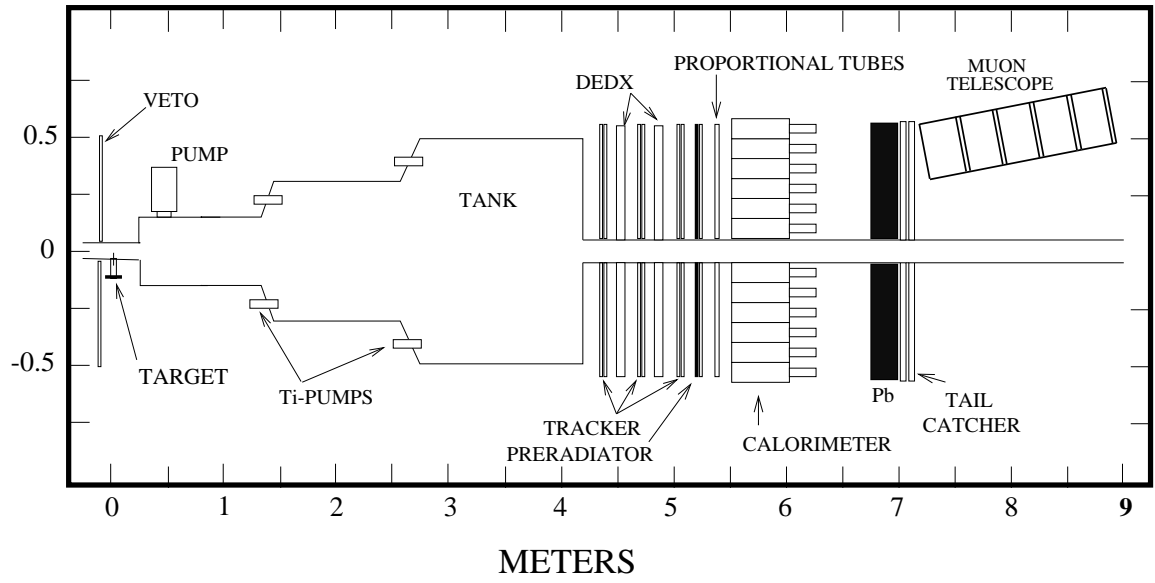}}
\vspace{0.5cm}
\caption{Schematic of the APEX detector.}
\label{detector_fig}
\end{figure}

\begin{figure}
\epsfxsize6.in
\centerline{\epsffile{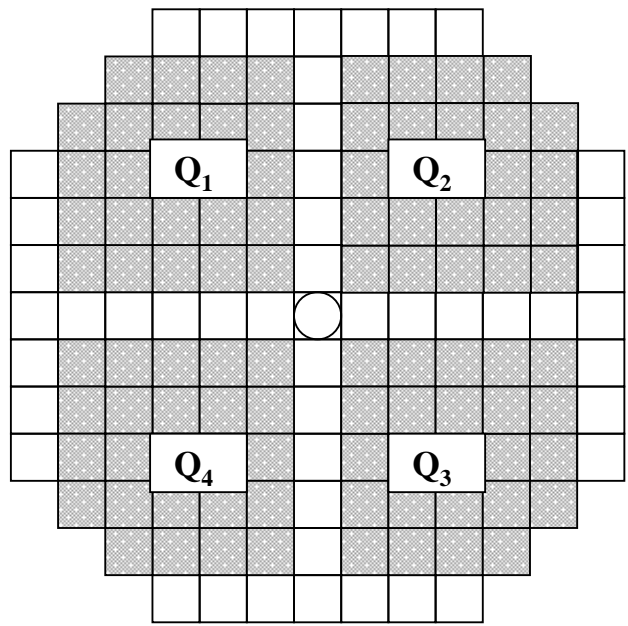}}
\caption{The four calorimeter quadrants used for triggering.}
\label{trigger_quads}
\end{figure}

The APEX detector was located in a 15.9~m straight section of the Fermilab
Antiproton Accumulator ring, which has a circumference of 474~m. The detector
was designed to identify $\overline{p}$ decays within a short 3.7~m long
section of the ring, and was optimized to detect a single energetic charged
track (electron or muon), originating from the beam, and accompanied by one
or more neutral pions or photons. A schematic of the detector is shown in
Fig.~\ref{detector_fig}. In the following we give a brief description of the
main detector components relevant to the analysis described in this paper. A
more detailed description can be found in Ref.~\cite{nim}. The APEX detector
consisted of the following:
\begin{description}
\item{(i)} A 3.7~m long evacuated decay tank operated at $10^{-11}$~Torr. 
The vacuum was maintained by a roughing pump, 
an ion diffusion pump, and twelve titanium sublimation filaments. 
The tank was designed to minimize the residual gas within the 
fiducial volume of the experiment, and hence minimize backgrounds 
from beam--gas interactions. The downstream section of the tank 
consists of a 96~cm diameter cylindrical shell supporting 
a 1.2~mm thick (0.073 radiation lengths) stainless steel window.
\item{(ii)} A movable tungsten wire target at the upstream end of the tank 
($z$ = 0 where, in the APEX coordinate system, $z$ is measured in the 
direction of the antiproton beam). 
The target could be inserted into the beam halo to produce a localized 
source of particles for aligning and calibrating the detector. 
\item{(iii)} An upstream veto system consisting of horizontal and vertical 
scintillation counters arranged 
around the 10~cm diameter beam pipe immediately upstream of 
the tank and target. The counters covered a \mbox{$1 \times 1$~m$^2$} 
area normal to the 
beam direction, and were used to reject tracks from upstream interactions.
\item{(iv)} Three planes of horizontal and three 
planes of vertical scintillation counters downstream of the tank. 
Each plane consisted of two $50 \times 100 \times 1.27$~cm$^3$ counters. 
The last planes of horizontal and vertical counters 
were downstream of a 2.3 radiation length lead wall, providing a preradiator 
to aid in identifying electrons and photons. The remaining counter planes 
were upstream of the lead, and provided pulse height information used to 
count the number of charged particles in an event ($dE/dx$ counters). 
The measured pulse height distribution arising from the passage of 
minimum ionizing particles (MIPs) through the $dE/dx$ counters 
is well described by a 
GEANT Monte Carlo simulation in which the yield is 37 photoelectrons per MIP. 
The predicted efficiency for a minimum ionizing particle to produce a counter 
signal in excess of a 0.5~MIP threshold is 97\%.
\item{(v)} A lead--scintillator sampling electromagnetic 
calorimeter~\cite{ref:fcal} constructed from 144 rectangular 
$10 \times 10$~cm$^2$ modules that are 14.7 radiation lengths deep. 
The modules were arranged in a $13 \times 13$ array with 6 modules 
removed from each of the 
four corners, and the central module removed to allow passage of the 
beam pipe (Fig.~\ref{trigger_quads}). 
Each module was read out with one 5.1~cm diameter photomultiplier. 
An LED pulser system enabled the stability of the photomultiplier 
gains for each module to be monitored. 
The calorimeter was calibrated by measuring the response to 
minimum ionizing tracks and reconstructing $\pi^0 \rightarrow \gamma\gamma$ 
and $\eta \rightarrow \gamma\gamma$ mass peaks 
using data samples recorded with the calibration target inserted into the 
beam halo. 
The measured mass peaks have 
fractional rms widths given by
$\sigma_m$/m~$\sim 0.25$ for cluster pairs in the energy range of
interest for the experiment. This 
mass resolution is dominated by the energy resolution of the calorimeter.
\item{(vi)} A tail catcher (TC) downstream of the calorimeter 
consisting of a 20~cm deep lead wall followed by planes of horizontal and
vertical scintillation counters.
\item{(vii)} A limited-acceptance muon telescope (MT), 10
nuclear interaction lengths deep, located downstream of the TC, 
and aligned to point towards the center of the decay tank. 
The MT consists of a sandwich of five iron plates and five 
$30 \times 30$~cm$^2$ scintillation counters. The MT was used to 
identify charged particles that penetrate through several interaction 
lengths of material (muon candidates).
\item{(viii)} A tracking system consisting of three planes of horizontal 
and three planes of vertical 2~mm diameter 
scintillating fibers downstream of the tank and upstream of the preradiator 
lead. The fibers were measured to yield about 9 photoelectrons for a 
traversing minimum ionizing particle, and 
were read out using 12-stage Hamamatsu multianode photomultipliers.
These detectors provided three space points 
along the track trajectory with typical residuals of 620~$\mu$m in the 
directions transverse to the beam direction. This resolution enabled the 
origin of tracks emerging from the decay tank to be reconstructed 
with an rms precision given by $\sigma_z = 12$~cm. 
The average single--hit efficiency for a minimum ionizing particle 
passing through a tracking plane was 85\%. 
The measured track reconstruction efficiency, for a sample of events 
that are consistent with 
having one minimum ionizing particle passing through the $dE/dx$ planes 
and the MT, is $(90 \pm 7)$\%.
\end{description}

\section{Data taking \protect\label{data}}

The APEX experiment took data during the April-July 1995 Fermilab Tevatron
collider running period at times when there were typically $10^{12}$ 
antiprotons circulating in the Accumulator ring with a central $\overline{p}$
momentum of $8.90 \pm 0.01$~GeV/c ($\gamma = 9.54 \pm 0.01$). APEX ran
parasitically to the Tevatron collider program. Due to increased background
interactions when antiprotons were being stacked in the Accumulator, data were
only taken during periods when stacking was not taking place. Such times
occurred during the short period (typically 90~minutes) before daily injection 
of antiprotons into the Tevatron, or at times when the collider complex was
not running due to maintenance or repairs. A measure of the sensitivity of the
APEX data sample is given by:
\begin{equation}
S \;\equiv\; \frac{1}{\gamma} \int{N_{\overline{p}}\,(t)\,dt}
\;=\; (3.31 \pm 0.03) \times 10^{9}\;\;yr,
\end{equation}
where N$_{\overline{p}}\,(t)$ is the number of circulating antiprotons at time
$t$, the integral is over the live-time of the experiment, and the uncertainty
arises from the precision with which the time dependent beam current was
recorded. Hence, if $\tau_{\,\overline{p}} = 3.31 \times 10^{9}$~years then on
average 0.63 antiproton decays would have occurred somewhere within the
Accumulator during the live-time of the experiment.

Energetic particles passing through the detector during Accumulator operation
predominantly arise from interactions of the $\overline{p}$ beam with the
residual gas in the decay tank or with material surrounding the beam. To
suppress these backgrounds, and select candidate \mbox{$\overline{p}
\rightarrow \mu^{-} X$} and \mbox{$\overline{p} \rightarrow e^{-} X$} decays,
signals from the calorimeter and the scintillation counters were used to form
muon and electron triggers.

\subsection{Muon trigger and data sample \protect\label{mutrig}}

To search for muonic $\overline{p}$ decays, data were recorded with a muon
trigger that required a coincidence between at least two of the five MT
scintillation counters. These triggers were eliminated if they were in
coincidence with a signal in one or more of the upstream veto counters
indicating the presence of an interaction upstream of the decay tank. This
loose trigger resulted in 1.2 $\times 10^{6}$ events being recorded with a
typical trigger rate of $\sim 1$~Hz. These events predominantly arise from
background interactions in which the coincident MT counter signals are
produced by traversing muons coming from charged pion decays, or by hadronic
showers not contained in the calorimeter and TC.

\subsection{Electron trigger and data samples \protect\label{etrig}}

To search for electronic $\overline{p}$ decays, data were recorded with three
complementary triggers: 
(i) An $E_{balance}$ trigger designed to select events consistent with a
two-body $\overline{p}$ decay that results in the deposition of energy in two
diagonally opposite quadrants of the calorimeter. The calorimeter cell
groupings into four trigger quadrants are shown in Fig.~\ref{trigger_quads}.
The trigger required the summed signals in each of two diagonally opposite
quadrants to exceed a threshold that was set equal to about 2~GeV; 
(ii) An $E_{total}$ trigger designed to select $\overline{p}$ decays in which
all of the decay products deposit approximately all of their energy within the
calorimeter. This would result in a total calorimeter energy deposition
$E_{total}$ close to the beam energy. The trigger therefore required the
summed signals from all of the calorimeter cells to exceed a threshold that
was set equal to about 7~GeV; and 
(iii) An $electron$ trigger which required signals in the scintillators of the
first two $dE/dx$ planes consistent with the passage of a MIP, a signal in the
corresponding preradiator scintillators exceeding about 1.5 MIPs, and a signal
in the corresponding calorimeter quadrant exceeding a summed energy threshold
set equal to about 5~GeV. 

All three triggers were combined with a signal from the upstream veto
counters, used in anti-coincidence to suppress events from upstream
interactions. The typical data taking rates from background events satisfying
the $E_{balance}$, $E_{total}$, and $electron$ trigger requirements were 
respectively 3~Hz, 9~Hz, and 8~Hz when there were $10^{12}$ antiprotons 
circulating in the accumulator. The $E_{balance}$, $E_{total}$, and $electron$
triggers resulted in a data sample of 37.8~million events recorded to tape for
further analysis.

\subsection{Calibration triggers \protect\label{caltrig}}

A number of additional triggers were used, either during normal data taking or
during special running periods, to collect data for monitoring and calibrating
the detector. Three of these triggers provided calibration data samples that
were specially important for the analysis: 
(1) A very low threshold rate limited calorimeter trigger ($E_{total} >
250$~MeV) that recorded events with minimal bias continuously during normal
running. Dedicated minimal bias data samples were also taken during extended
periods when there was a low intensity $\overline{p}$ current (corresponding
to $\sim 10^{11}$ stored antiprotons) in the Accumulator; 
(2) A traversing charged track trigger that required a coincidence between the
upstream veto counters and the TC counters. This trigger provided a sample of
MIPs passing through the calorimeter; and 
(3) Low threshold ($E_{total} > 5$~GeV) and high threshold ($E_{total} >
7$~GeV) calorimeter triggers used during special periods during which the
calibration target was inserted into the beam halo. This provided tracks
coming from a known origin which facilitated the relative alignment of the
tracking planes. These data also provided a sample of localized interactions
which enabled the reconstruction of the $\pi^0 \rightarrow \gamma\gamma$ and
$\eta \rightarrow \gamma\gamma$ peaks used to define the calorimeter
calibration.

\section{Search for antiproton decay with a final state muon
\protect\label{musearch}}

We begin with the sample of 1.2 $\times 10^{6}$ events recorded with the muon
trigger. Offline, after final calibration of the scintillation counters, the
upstream veto counter requirement was re-applied using a more stringent
threshold. This reduced the data sample to 1.1 $\times 10^{6}$ events. The
remaining backgrounds can be further suppressed by requiring that (a) the
scintillation counter and scintillating fiber tracker signals are consistent
with the presence of a minimum ionizing charged track that points back to the
beam trajectory within the decay tank and that points forward to a pattern of
minimum ionizing energy depositions within the MT counters, and (b) an event
topology and kinematics that is consistent with the decay
$\overline{p}$~$\rightarrow \mu^- + X$. In the following, we consider first
the simplest decay channel $\overline{p}$~$\rightarrow \mu^- \gamma$, and then
describe the search for more complicated decay modes.

\subsection{$\overline{p} \rightarrow \mu^- \gamma$ event selection
\protect\label{mugamma}}

\begin{figure}
\vspace{-2.2cm}
\epsfxsize6.in
\centerline{\epsffile{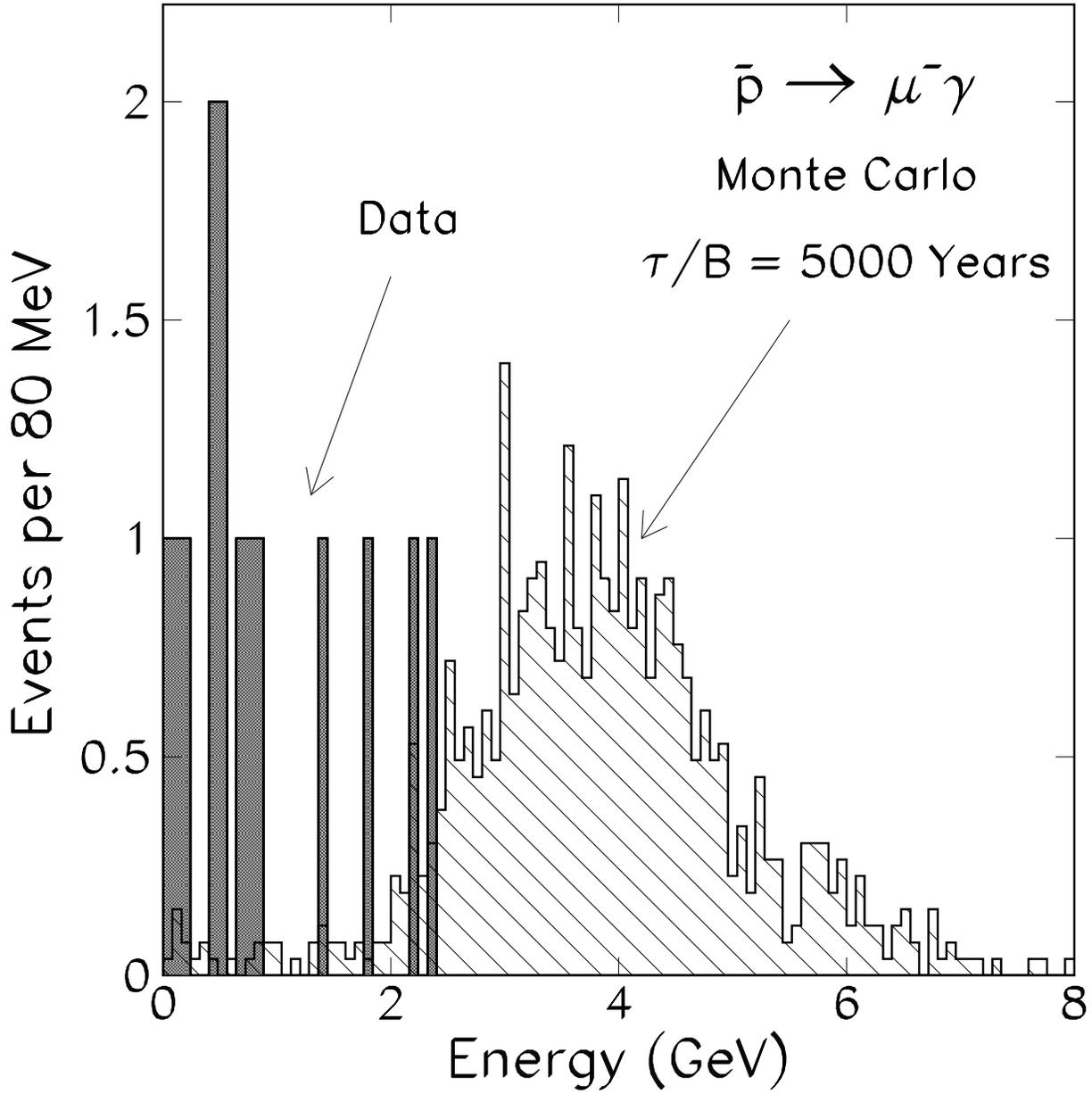}}
\vspace{-2.2cm}
\caption{Distribution of neutral cluster energies for the 
14 events that pass the $\overline{p} \rightarrow \mu^- \gamma$ 
selection criteria described in the text (open histogram) compared 
with the predicted distribution for a signal corresponding to 
$\tau_{\,\overline{p}}/B = 5000$~yr (hatched histogram).}
\label{mugam_fig}
\end{figure}

To select events consistent with the decay $\overline{p}$~$\rightarrow \mu^-
\gamma$ we begin by requiring that: 
(i) at least four of the five MT counters be above threshold ($4.2 \times 10^4$
events), and 
(ii) there be one and only one reconstructed track that extrapolates to the MT
counters within the expected uncertainty due to multiple scattering, and also
extrapolates back to the beam orbit with a point of closest approach within the
fiducial volume of the decay tank ($0 < z < 375$~cm) and with an impact
parameter~\cite{nim} less than 1~cm (416 events). Since a $\overline{p}
\rightarrow \mu^{-}\gamma$ decay would result in events in which an energetic
photon is produced coplanar with the muon (i.e. traveling within the plane
defined by the muon and the incoming beam direction), we next require that 
(iii) there be at least one neutral cluster~\cite{nim} in the calorimeter (209
events) that is coplanar with the muon candidate ($\pm 5^\circ$). The data
sample is reduced to 14 events. The observed neutral cluster energy
distribution for these events is compared in Fig.~\ref{mugam_fig} with the
predicted distribution obtained using the GEANT~\cite{geant} simulation
described below, and corresponding to $\overline{p} \rightarrow \mu^-\gamma$
decay with a lifetime $\tau_{\,\overline{p}}$/B$(\overline{p} \rightarrow
\mu^-\gamma) = 5000$~yr. The observed distribution peaks at low cluster
energies, with a tail extending to approximately 2~GeV. In contrast, the
predicted distribution for $\overline{p} \rightarrow \mu^-\gamma$ decays peaks
at about 3.5~GeV, with only 4.4\% of the simulated events having cluster
energies less than 2~GeV. We conclude that there is no evidence for a signal. 

\subsection{Limit on $\overline{p}$~$\rightarrow \mu^- \gamma$ decays
\protect\label{mugammalimit}}

To set a limit on $\tau_{\,\overline{p}}$/B($\overline{p} \rightarrow
\mu^-\gamma$) the remaining background events are eliminated by requiring 
that the neutral cluster energy exceed a minimum value $E_{min}$, where we
choose $E_{min} = 3$~GeV. We note that although the choice of $E_{min}$ is
somewhat arbitrary, the final limit that we obtain on the decay $\overline{p}
\rightarrow \mu^-\gamma$ is insensitive to small changes in $E_{min}$. After
setting $E_{min} = 3$~GeV, the resulting limit on
$\tau_{\,\overline{p}}$/B($\overline{p} \rightarrow \mu^-\gamma$) is given in
years by: 
\begin{equation}
\tau_{\,\overline{p}}/B(\overline{p} \rightarrow \mu^-\gamma) \; > \;
- \; \frac{1}{ln(1-N_{max}/\epsilon S)},
\end{equation}
where $N_{max}$ = 2.3 is the 90\% C.L. upper limit on the observation of $N =
0$ events, and $\epsilon$ is the fraction of decays taking place uniformly
around the accumulator ring that would pass the trigger and event selection
requirements. If $(N_{max}/\epsilon S) << 1$, to a good approximation :
\begin{equation}
\tau_{\,\overline{p}}/B(\overline{p} \rightarrow \mu^-\gamma) \; > \;
\frac{\epsilon}{\gamma} \; \frac{1}{N_{max}} \; \int{N_{\overline{p}}\,(t)\,dt}
\; = \; (3.31 \pm 0.03) \times 10^9 \; \frac{\epsilon}{N_{max}}.
\end{equation}
To take account of $\sigma_{\epsilon}$, the systematic uncertainty on
$\epsilon$, we use the prescription given in Ref.~\cite{cousins}, giving at
90\% C.L.
\begin{equation}
N _{max} \; = \; \mu_{max} \times (1 + \mu_{max}~\sigma_r^2/2),
\end{equation}
where $\sigma_r \equiv \sigma_\epsilon/\epsilon$, and $\mu_{max}$ is the upper
limit corresponding to the observation of $N$ events ($\mu_{max} = 2.3$ for $N
= 0$).

The GEANT Monte Carlo program has been used to simulate the detector response
and calculate $\epsilon$. The detector simulation includes a full description
of the detector geometry, and correctly describes the calorimeter, tracker, and
scintillation counter responses (dE/dx, MT) measured using calibration data,
together with the measured performance of the muon trigger. Further details can
be found in Ref.~\cite{nim}. The efficiency $\epsilon$ was obtained by
generating $10^5 \; \overline{p} \rightarrow \mu^{-}\gamma$ decays uniformly
along the beam orbit within the decay tank, and applying the trigger and data
selection requirements to the simulated events. The geometrical efficiency of
the detector and trigger is negligible for decays occurring outside of the
tank. We obtain $\epsilon = (3.7 \pm 0.9) \times 10^{-5}$, where the
uncertainty on $\epsilon$ arises from the systematic uncertainties on the muon
trigger counter threshold calibration, the calorimeter energy scale, and the
track reconstruction efficiency. The contributions to $\sigma_{r}$ arising
from the uncertainties on the muon trigger counter threshold calibration and
the calorimeter energy scale are, respectively, $\pm18$\% and $\pm15$\%. These
uncertainties have been evaluated by analyzing GEANT Monte Carlo $\overline{p}
\rightarrow \mu^{-}\gamma$ samples in which the simulated trigger calibration
and the calorimeter energy scale have been changed by $\pm1$~standard
deviation. The overall systematic uncertainty on $\epsilon$ was calculated by
adding these contributions in quadrature with the uncertainty on the track
reconstruction efficiency ($\pm7$\%). Inserting the calculated $\epsilon$ and
$\sigma_\epsilon$ into Eqs.~3 and 5 we obtain the result~\cite{muonprl} 
$\tau_{\,\overline{p}}/B(\overline{p} \rightarrow \mu^-\gamma) \; > \;$ $5
\times 10^4$~yr (90\% C.L.).

Finally, we note that although our limit is insensitive to the exact value of
$E_{min}$, the dependence of the final result on the position of this last
cut is somewhat unsatisfying. Hence, to verify our result, we have also used
the more complicated statistical method described in Ref.~\cite{apex_note}. In
this alternative statistical analysis all of the events in the final
($E_{min}$) distribution are retained, and use is made of how ``signal--like"
the position of each event is in the final distribution. Using this method we 
obtain a limit on $\tau_{\,\overline{p}}/B(\overline{p} \rightarrow
\mu^-\gamma)$ that is consistent with our quoted result.

\subsection{Search for $\overline{p}$~$\rightarrow \mu^- \pi^0$, 
$\overline{p}$~$\rightarrow \mu^- \eta$, 
and $\overline{p}$~$\rightarrow \mu^- \gamma\gamma$ decays 
\protect\label{mugamgam}}

Now consider the two--body decays $\overline{p} \rightarrow \mu^-\pi^0$ and
$\overline{p} \rightarrow \mu^-\eta$ ($\pi^0 ,\; \eta \rightarrow
\gamma\gamma$), and the three--body decay $\overline{p} \rightarrow
\mu^-\gamma\gamma$. These decays would result in events with one or two neutral
clusters observed in the calorimeter, where the one--cluster events occur when
the two photon--showers are not spatially resolved in the calorimeter or when
one of the photons is outside of the calorimeter acceptance. In addition, the
$\eta$ may also decay into more complicated final states producing further
clusters in the calorimeter. To optimize the search for $\mu^-\pi^0$ and
$\mu^-\eta$ final states we divide the 209 events described previously that
have a muon candidate plus one or more neutral clusters into two subsamples,
namely a one--cluster sample containing 104 events, and a multi--cluster sample
containing 105 events. The position of the cluster in the one--cluster sample
is required to be coplanar with the muon ($\pm 10^\circ$). The position of the
cluster--pair formed by the energy-weighted centroid of the two highest energy
clusters in the multi--cluster events is also required to be coplanar with the
muon ($\pm 10^\circ$). These requirements reduce the samples to 13 one-cluster
and 10 multi-cluster events. Note that we expect the multi-cluster events to
contain 2 photons if they are genuine $\overline{p} \rightarrow \mu^-\pi^0,
\mu^-\eta$, or $\mu^-\gamma\gamma$ decays. In most cases at least one of the
two photons would be expected to convert in the preradiator lead. We therefore
require that the preradiator counters in the quadrant opposite the MT have a
signal exceeding a threshold of 0.5~MIPs. Only 3 multi-cluster events satisfy
this requirement. Hence we are left with 16 events with $\ge$~1 cluster ($\mu
\;+ \ge 1$~cluster events) for further analysis.

\begin{figure}
\vspace{-2.2cm}
\epsfxsize6.in
\centerline{\epsffile{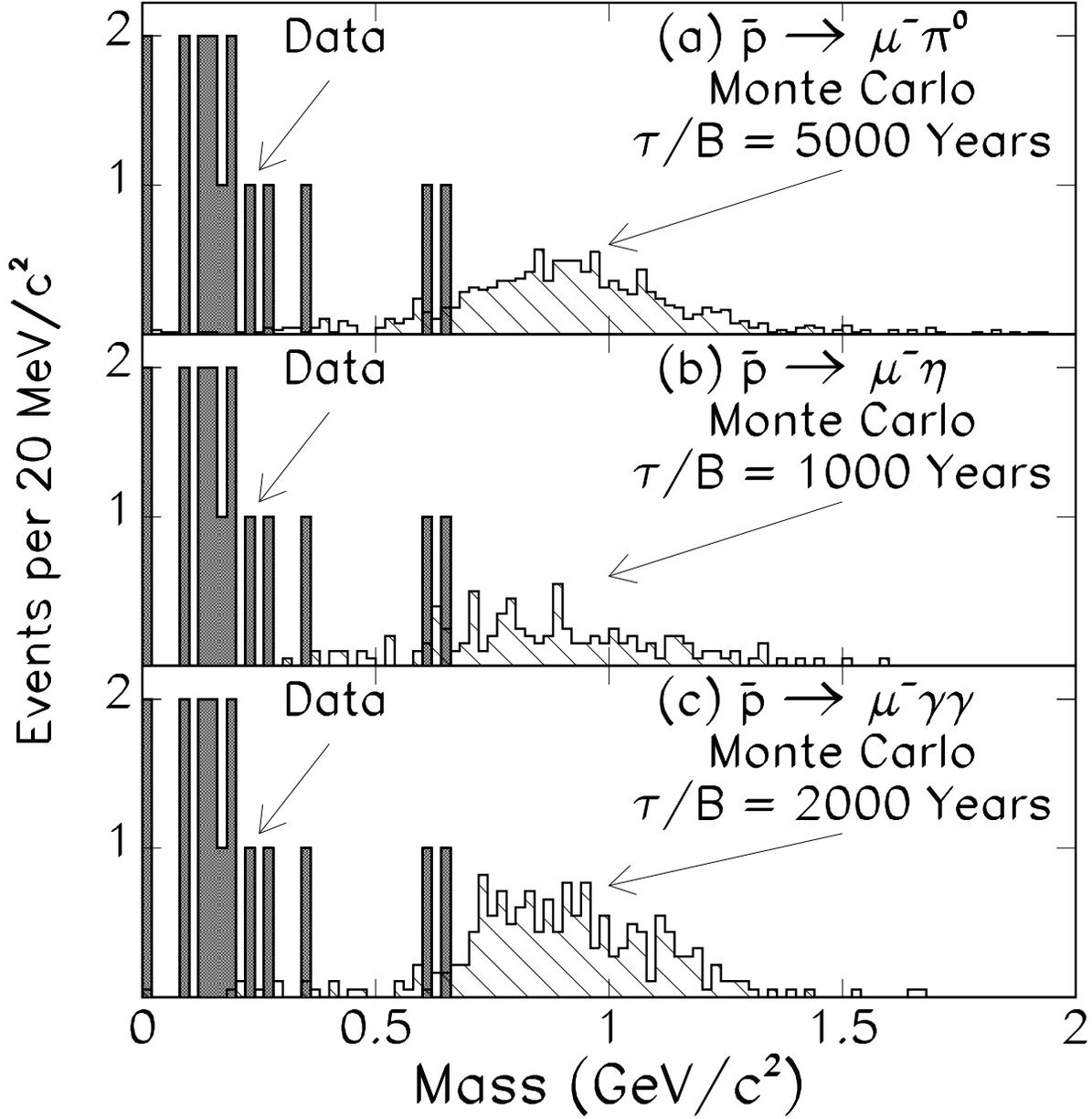}}
\vspace{-2.2cm}
\caption{Distribution of event masses for 
the 16 events that pass the selection criteria described in the text 
(open histograms) compared with predictions from a GEANT simulation 
(hatched histograms) for the decays 
(a) $\overline{p} \rightarrow \mu^-\pi^0$, 
(b) $\overline{p} \rightarrow \mu^-\eta$, and 
(c) $\overline{p} \rightarrow \mu^-\gamma\gamma$. The two entries in the 
lowest bin are events with calorimeter energy less than the $\pi^0$ 
rest energy. The predicted signal distributions are normalized to 
correspond to $\tau_{\,\overline{p}}/B = 5000$~yr for the $\mu^-\pi^0$ mode, 
1000 yr for the $\mu^-\eta$ mode, and 2000~yr for the 
$\mu^-\gamma\gamma$ mode.}
\label{mass_fig}
\end{figure}

Under the hypothesis that the observed ($\mu\gamma$) and ($\mu\gamma\gamma$)
systems arise from the decay of a beam particle, the mass of the parent
particle can be computed from the measured muon direction and the directions
and energies of the neutral clusters, using the constraint that the vector sum
of the momentum components of the daughter particles transverse to the beam
direction is zero. The resulting mass distribution for the remaining 16 ($\mu
\;+ \ge 1$~cluster) events is compared in Fig.~\ref{mass_fig} with predictions
from the GEANT simulation for the decays (a) $\overline{p} \rightarrow
\mu^-\pi^0$, (b) $\overline{p} \rightarrow \mu^-\eta$, and (c) $\overline{p}
\rightarrow \mu^-\gamma\gamma$. The observed mass distribution peaks at low
masses with a tail extending to approximately 0.7~GeV/c$^2$. In contrast to
this, the simulated signal distributions peak at the $\overline{p}$ mass, with
only 17\% (23\%) [13\%] of the simulated $\mu^-\pi^0$ ($\mu^-\eta$)
[$\mu^-\gamma\gamma$] decays resulting in reconstructed masses less than
0.7~GeV/c$^2$. We conclude that there is no evidence for a signal. 

To set limits on $\tau_{\,\overline{p}}/B$ we eliminate the remaining
background by requiring that the reconstructed mass exceed a minimum value
$m_{min}$, and choose $m_{min} = 0.75$~GeV/c$^2$. We note that although the
choice of $m_{min}$ is somewhat arbitrary, the final limits that we obtain on
the decays $\overline{p} \rightarrow \mu^-\pi^0$, $\mu^-\eta$, and
$\mu^-\gamma\gamma$ are insensitive to small changes in $m_{min}$. The
calculated overall efficiencies $\epsilon$ for these decays to pass our trigger
and analysis requirements are $(3.6 \pm 0.9) \times 10^{-5}$ for the
$\mu^-\pi^0$ mode, $(6.1 \pm 1.9) \times 10^{-6}$ for the $\mu^-\eta$ mode, and
$(1.8 \pm 0.5) \times 10^{-5}$ for the $\mu^-\gamma\gamma$ mode. Substituting
the values into Eqs.~3 and 5 yields the limits
$\tau_{\,\overline{p}}/B(\overline{p} \rightarrow \mu^-\pi^0) \; > \;$ $5
\times 10^4$~yr, $\tau_{\,\overline{p}}/B(\overline{p} \rightarrow \mu^-\eta)
\; > \;$ $8 \times 10^3$~yr, and $\tau_{\,\overline{p}}/B(\overline{p}
\rightarrow \mu^-\gamma\gamma) \; > \;$ $2 \times 10^4$~yr at 90\% C.L.

Finally, to verify our result, we have removed the last cut on our final 
(mass) distribution, and used the more complicated statistical method
described in Ref.~\cite{apex_note}. The resulting limits on the $\mu^-\pi^0$,
$\mu^-\eta$, and $\mu^-\gamma\gamma$ decay modes are consistent with our
quoted results.

\subsection{Other $\overline{p}$~$\rightarrow \mu^- + X$ decays 
\protect\label{muother}}

Now consider the other possible two--body muonic decay modes of the antiproton,
namely decays into the final states $\mu^-K^0_S$, $\mu^-K^0_L$, $\mu^-\rho^0$,
and $\mu^-\omega$. Detailed GEANT simulations have been made for these decay
modes. The calculated geometrical acceptances for the $\mu^-\rho^0$ and
$\mu^-\omega$ modes are very small. We therefore restrict ourselves to the
$\mu^-K^0_S$ and $\mu^-K^0_L$ modes. We have not made an explicit event
selection for the $\mu^-K^0_S$ and $\mu^-K^0_L$ modes. However, the GEANT
simulations predict the fraction of these decays that would have passed our
$\overline{p} \rightarrow \mu^-\gamma$ and $\overline{p} \rightarrow
\mu^-\pi^0$ selection requirements. We can use these calculated efficiencies
to set an upper limit on the $\mu^-K^0_S$ and $\mu^-K^0_L$ modes. The
calculated efficiencies $\epsilon$ for these decays to satisfy the trigger and
either the $\mu^-\gamma$ or $\mu^-\pi^0$ search criteria described previously
are $(3.3 \pm 1.0) \times 10^{-6}$ for the $\mu^-K^0_S$ mode, and $(5.0 \pm
1.5) \times 10^{-6}$ for the $\mu^-K^0_L$ mode. Substituting these values into
Eqs.~3 and 5 yields the limits $\tau_{\,\overline{p}}/B(\overline{p}
\rightarrow \mu^-K^0_S) \; > \;$ $4 \times 10^3$~yr and
$\tau_{\,\overline{p}}/B(\overline{p} \rightarrow \mu^-K^0_L) \; > \;$ $7
\times 10^3$~yr at 90\% C.L. To verify these results we have removed the last
cut on our final kinematic distribution and used the more complicated
statistical method described in Ref.~\cite{apex_note}. The resulting limits on
the $\mu^-K^0_S$ and $\mu^-K^0_L$ decay modes are consistent with our quoted
results.

\section{Search for antiproton decay with a final state electron
\protect\label{esearch}}

We begin with the sample of $37.8 \times 10^6$ events recorded with the 
$E_{balance}$, $E_{total}$, and $electron$ triggers described in
Sec.~\ref{etrig}. 
Offline, after final calibration of the scintillation counters, a
re-application of the upstream veto counter requirement reduces this sample to
$34.2 \times 10^6$ events. The remaining backgrounds can be suppressed by
requiring that the signals from the tracker, calorimeter, and scintillation
counters are consistent with events arising from $\overline{p} \rightarrow e^-
+ X$ decays. In the following we first consider the simplest decay channel
$\overline{p} \rightarrow e^- \gamma$, and then describe the search for more
complicated decay modes. 

\subsection{$\overline{p} \rightarrow e^{-} \gamma$ event selection
\protect\label{egamma}}

To select events consistent with the decay \mbox{$\overline{p} \rightarrow
e^{-} \gamma$} we begin by requiring:
\begin{description}
\item{(1)} A pattern of hits in the $dE/dx$ counters consistent
with the presence of one, and only one, charged track
($16.9 \times 10^6$ events).
\item{(2)} A pulse height in excess of 1.5~MIPs in the preradiator
counters ($9.6 \times 10^6$ events). This is a first, very loose, 
electron candidate requirement. 
\item{(3)} No signal in excess of 0.6~MIPs in both a horizontal 
and vertical TC counter ($7.6 \times 10^6$ events). 
Note that electrons and photons 
are expected to produce electromagnetic showers that are fully 
contained within the calorimeter and the TC lead wall, 
and do not penetrate into the TC.
\item{(4)} Either 2 or 3 clusters reconstructed in the
calorimeter ($2.5 \times 10^6$ events). A fully contained 
\mbox{$\overline{p} \rightarrow e^{-} \gamma$} 
event would normally result in two calorimeter 
clusters. However a third cluster can arise if, for example, 
a large angle photon is 
radiated from the electron as it passes through the preradiator. 
\item{(5)} More than 90\% of the total energy recorded in 
the calorimeter be outside of the cells immediately 
surrounding the beampipe ($1.3 \times 10^6$ events). This requirement 
suppresses backgrounds associated with beam halo outside of, but 
close to, the beampipe. The requirement also 
ensures that a significant fraction of the energy associated 
with the observed electromagnetic showers has not ``leaked" out of the 
calorimeter into the uninstrumented region within the beampipe. 
\end{description}
These very loose selection criteria were designed to reduce the sample to a
reasonable size before running the track reconstruction code. After track
reconstruction we impose the following requirements:
\begin{description}
\item{(6)} There must be one and only one reconstructed track that 
extrapolates to a calorimeter cluster ($1.0 \times 10^6$ events). 
\item{(7)} The reconstructed track must be consistent with originating
from within the fiducial volume of the decay tank. Specifically, 
the point of closest approach of the track to the beamline 
must be in the region $0<z<375$~cm ($2.7 \times 10^5$ events).
\item{(8)} The distance of closest approach of the reconstructed 
track to the beamline~\cite{nim} must be less than 1~cm 
($6.0 \times 10^4$ events).
\item{(9)} There must be a preradiator signal that exceeds 4~MIPs 
($2.2 \times 10^4$ events). This requirement was based on a study of 
the preradiator signals associated with the 
$\pi^0 \rightarrow \gamma\gamma$ calibration data samples. 
\item{(10)} The electron 
candidate must be coplanar ($\Delta\phi > 3.05$~radians) with the beamline and 
the highest energy neutral cluster (467 events). 
\end{description}

\begin{figure}
\vspace{-2.2cm}
\epsfxsize6.in
\centerline{\epsffile{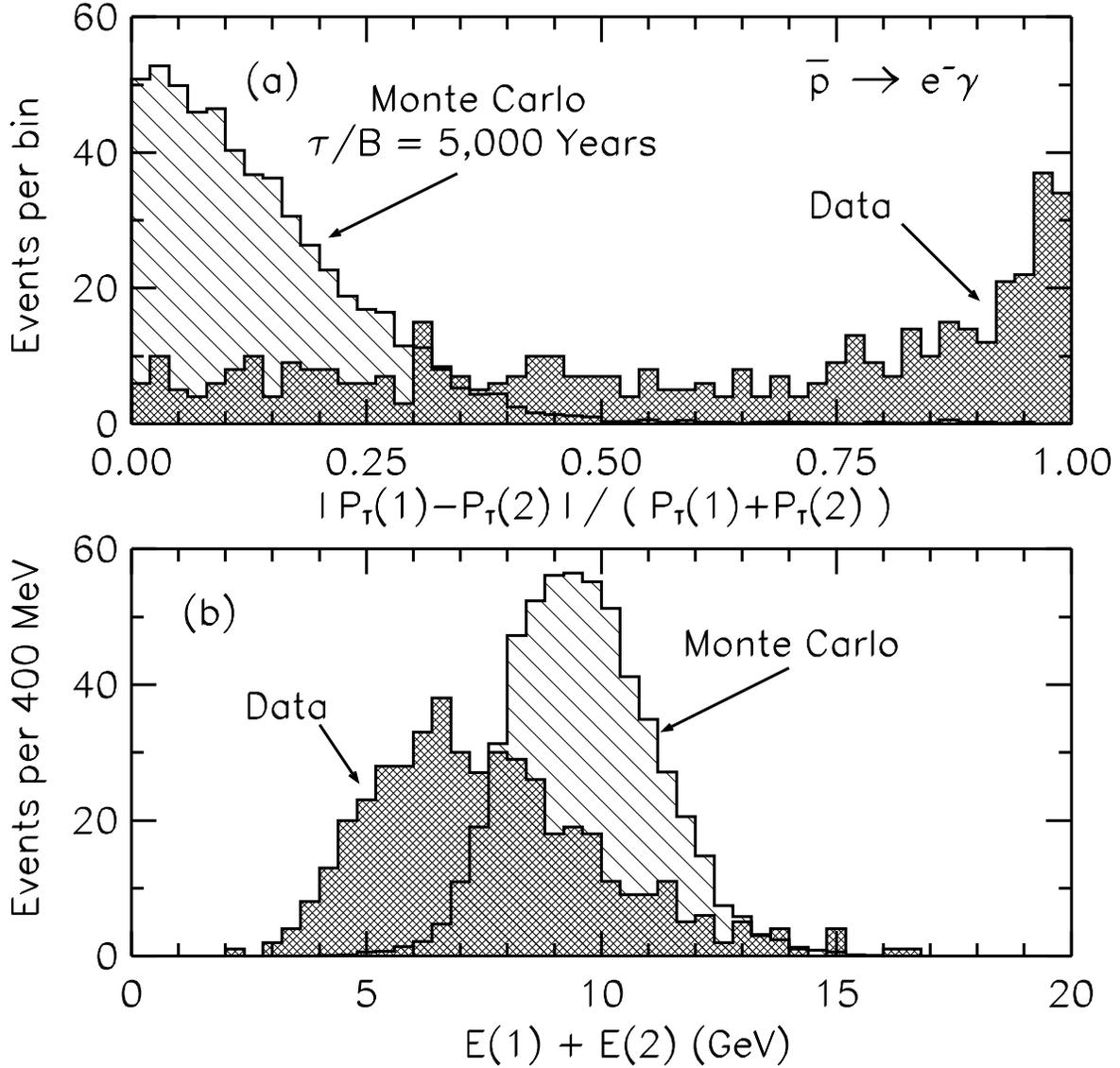}}
\vspace{-2.2cm}
\caption{Kinematic distributions for the 467 
$\overline{p} \rightarrow e^-\gamma$ candidate events prior to the 
kinematic analysis described in the text. The observed distributions 
are compared with GEANT Monte Carlo predictions for 
$\overline{p} \rightarrow e^-\gamma$ decays normalized to correspond to 
a lifetime of 5000~years, and shown for (a) a measure of the 
transverse momentum balance in the event, where $P_T(1)$ and $P_T(2)$ 
are the transverse momenta of the two highest energy clusters in 
the event, and 
(b) the sum of the energies of the two highest energy clusters 
in the event.}
\label{egam_fig}
\end{figure}

Figures~\ref{egam_fig}a and \ref{egam_fig}b compare, respectively, the
transverse--momentum--balance and total--energy distributions for these 467
events with predictions from a GEANT simulation of \mbox{$\overline{p}
\rightarrow e^{-} \gamma$} decays. The observed distributions have peaks
corresponding to large transverse--momentum imbalances and low total energies. 
In contrast to this, the simulated signal distributions peak at small
transverse--momentum imbalance and at total energies corresponding to the beam
energy. The backgrounds can therefore be further suppressed by requiring that
candidate events are kinematically consistent with \mbox{$\overline{p}
\rightarrow e^{-} \gamma$} decays.

\subsection{$\overline{p} \rightarrow e^{-} \gamma$ kinematics and a limit on 
$\tau/B$ \protect\label{egammakine}}

\begin{figure}
\vspace{-2.2cm}
\epsfxsize6.in
\centerline{\epsffile{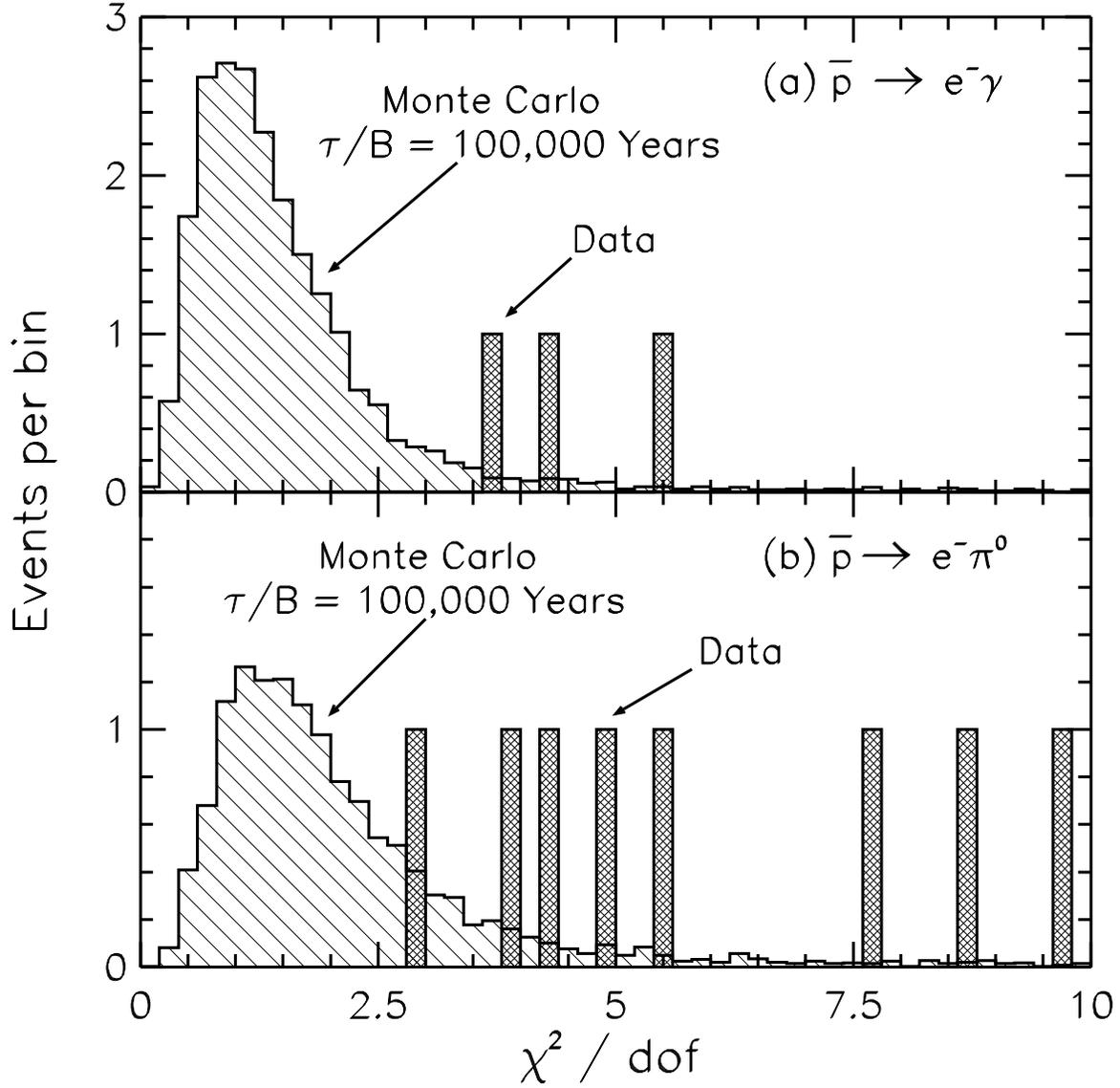}}
\vspace{-2.2cm}
\caption{Distribution of $\chi^2$ per degree of freedom for 
(a) the 3 $\overline{p} \rightarrow e^-\gamma$ candidate events surviving the 
kinematic fitting described in the text, compared with the 
GEANT Monte Carlo predictions for 
$\overline{p} \rightarrow e^-\gamma$ decays normalized to correspond to 
a lifetime of 100,000~years, and 
(b) the 8 $\overline{p} \rightarrow e^-\pi^0$ candidate events surviving the
kinematic fitting described in the text, compared with the
GEANT Monte Carlo predictions for
$\overline{p} \rightarrow e^-\pi^0$ decays normalized to correspond to
a lifetime of 100,000~years.}
\label{chisq_fig}
\end{figure}

In general, a two-body decay in which the daughter particles have known masses
can be completely specified by 9 parameters (two momentum--vectors and a decay
point $x_0, y_0, z_0$). Consider \mbox{$\overline{p} \rightarrow e^{-}
\gamma$} decays in which the parent antiprotons are traveling along the
beamline with momenta $p_0$, energy $E_0$. In this case there are 6
constraints: (i) $x_0 = 0$, (ii) $y_0 = 0$, (iii) $E(e) + E(\gamma) = E_0$, 
(iv) $p_x(e) + p_x(\gamma) = 0$, (v) $p_y(e) + p_y(\gamma) = 0$, and (vi)
$p_z(e) + p_z(\gamma) = p_0$. Hence, to completely specify a 
\mbox{$\overline{p} \rightarrow e^{-} \gamma$} decay occurring within the APEX
decay tank, we need to specify the values of 3 parameters, which we choose to 
be $z_0$, $E(e)$, and the $x$--coordinate of the intersection of the electron
trajectory with the calorimeter $x_{cal}(e)$. In the following we will refer
to a specific set of values for these parameters as a decay hypothesis. For a
given decay hypothesis the intersections of the electron track with the 3
horizontal and 3 vertical tracking planes can be predicted, along with the
positions and energies of the calorimeter clusters associated with the
electron and photon. To quantify how well a given decay hypothesis describes
the observed locations of the hits in the tracker planes, and the observed
locations and energies of the clusters in the calorimeter, we define: 
\begin{equation}
\chi^2 \; \equiv \; \sum_{i=1}^{n}\; \left[ \frac{\Delta_i}{\sigma_{i}} 
\right]^2, 
\end{equation}
where the sum is over the 6 calorimeter measurements ( [x,y]--positions and 
energies for the electron and photon clusters) and ($n-6$) tracking
measurements. The $\Delta_i$ are the differences between the measured and
predicted values for the $i^{th}$ measurement, and the $\sigma_{i}$ are the
corresponding measurement uncertainties. Note that the track reconstruction
code requires that at least 4 of the 6 scintillating fiber planes record a hit
along the track, and hence the number of measurements used in the $\chi^2$ sum 
is in the range $10 \le n \le 12$. The best decay hypothesis for a given event
is found by finding the set of parameters ($z_0$, $E(e)$, and $x_{cal}(e)$)
that minimizes the $\chi^2$ per degree of freedom ($\chi^2 / dof = \chi^2 /
n$). In this fitting procedure, hypotheses are rejected if 
the $\chi^2 / dof > 10$. Only 3 of the 
467 events shown in Fig.~\ref{egam_fig}a have a best--fit decay hypothesis
that survives this loose kinematic requirement. In Fig.~\ref{chisq_fig}a the
distribution of the fitted values of $\chi^2 / dof$ for these events are
compared with the predicted GEANT Monte Carlo distribution for 
\mbox{$\overline{p} \rightarrow e^{-} \gamma$} decays that pass the trigger
and offline selection requirements. There is no evidence for a signal. The
observed events have values of $\chi^2 / dof$ larger than would be expected 
for typical \mbox{$\overline{p} \rightarrow e^{-} \gamma$} decays. 

To set a limit on $\tau / B$ we require that the $\chi^2 / dof$ for the
best--fit hypothesis is less than a maximum value, which we set equal to 2. No
events survive this requirement. The efficiency for \mbox{$\overline{p}
\rightarrow e^{-} \gamma$} decays to pass the trigger and offline selection
requirements has been evaluated using the GEANT Monte Carlo program to
generate $10^5$ simulated decays uniformly along the beamline within the decay
tank. We find $\epsilon = (5.2 \pm 1.4) \times 10^{-4}$, where the uncertainty 
on $\epsilon$ arises from the systematic uncertainties on the trigger 
calibration, calorimeter energy scale, track reconstruction efficiency, and
Monte Carlo statistics. The trigger and calorimeter scale uncertainties yield
contributions to $\sigma_{r}$ of $\pm 14$\% and $\pm 21$\%, respectively, and
have been evaluated by analyzing GEANT Monte Carlo $\overline{p} \rightarrow
e^{-}\gamma$ samples in which the simulated trigger and calorimeter scales have
been changed by $\pm1\,\sigma$. The overall systematic uncertainty on
$\epsilon$ was calculated by adding these contributions in quadrature with the
uncertainty on the track reconstruction efficiency ($\pm7$\%) and Monte Carlo
statistics ($\pm$1\%). Inserting the calculated $\epsilon$ and
$\sigma_\epsilon$ into Eqs.~3 and 5 we obtain the result 
$\tau_{\,\overline{p}}/B(\overline{p} \rightarrow e^-\gamma) \; > \;$ $7 \times
10^5$~yr (90\% C.L.). To verify this result we have removed the cut on the
$\chi^2 / dof$ distribution and used the more complicated statistical method
described in Ref.~\cite{apex_note}. The resulting limit on 
$\tau_{\,\overline{p}}/B(\overline{p} \rightarrow e^-\gamma)$ is consistent
with our quoted result.

\subsection{Search for $\overline{p} \rightarrow e^{-} \pi^{0}$ 
\protect\label{epiz}}

To select events consistent with the decay \mbox{$\overline{p} \rightarrow
e^{-} \pi^{0}$} we begin by applying the \mbox{$\overline{p} \rightarrow e^{-}
\gamma$} selection requirements (1)--(9) described in Sec.~\ref{egamma}. These
requirements select $2.2 \times 10^4$ events in which there is a candidate
electron together with 1 or 2 additional clusters in the calorimeter. We would
expect genuine \mbox{$\overline{p} \rightarrow e^{-} \pi^{0}$} decays to be
either reconstructed as an electron plus one additional cluster (if the two
photons from the $\pi^0$ decay are not spatially resolved in the calorimeter),
or as an electron plus two additional clusters (if the two photons are
resolved). If there are two additional clusters, we require the 
cluster--cluster mass to be less than 500~MeV/$c^2$ ($2.1 \times 10^4$ events).
It is convenient to analyze all the \mbox{$\overline{p} \rightarrow e^{-}
\pi^{0}$} candidate events as two--body decays. To do this we reduce the
electron plus two cluster events to an electron plus one cluster topology by 
combining the two neutral clusters. The combined cluster has an energy equal to
the sum of the energies of the component clusters, and is located in the
(x,y)--plane at a position given by the energy-weighted centroid of the
component clusters. We now require that the electron candidate is coplanar 
($\Delta \phi > 2.9$~radians) with the beamline and the neutral cluster (1091
events). Applying the two--body kinematics fitting procedure described in
Sec.~\ref{egammakine}, only 8 events have $\chi^2/dof < 10$. In
Fig.~\ref{chisq_fig} the distribution of the fitted values of $\chi^2 / dof$
for these events are compared with the predicted GEANT Monte Carlo distribution
for \mbox{$\overline{p} \rightarrow e^{-} \pi^{0}$} decays that pass the
trigger and offline selection requirements. There is no evidence for a signal.
The observed events have values of $\chi^2 / dof$ larger than would be expected
for \mbox{$\overline{p} \rightarrow e^{-} \pi^{0}$} decays.

To set a limit on $\tau / B$ we require that the $\chi^2 / dof$ for the
best--fit hypothesis is less than a maximum value, which we set equal to 2.5.
No events survive this requirement. The efficiency for \mbox{$\overline{p}
\rightarrow e^{-} \pi^0$} decays to pass the trigger and offline selection
requirements has been evaluated using the GEANT Monte Carlo to generate $10^5$
simulated decays uniformly along the beamline within the decay tank. We find
$\epsilon = (3.0 \pm 0.7) \times 10^{-4}$, where the uncertainty on $\epsilon$
arises from the systematic uncertainties on the trigger calibration,
calorimeter energy scale, track reconstruction efficiency, and Monte Carlo
statistics. Inserting the calculated $\epsilon$ and $\sigma_\epsilon$ into
Eqs.~3 and 5 we obtain the result $\tau_{\,\overline{p}}/B(\overline{p}
\rightarrow e^-\pi^0) \; > \;$ $4 \times 10^5$~yr (90\% C.L.). To verify this
result, we have removed the cut on the $\chi^2 / dof$ distribution, and used
the more complicated statistical method described in Ref.~\cite{apex_note}.
The resulting limit on $\tau_{\,\overline{p}}/B(\overline{p} \rightarrow
e^-\pi^0)$ is consistent with our quoted result. We have also searched for 
$\overline{p} \rightarrow e^-\pi^0$ decays by employing an analysis in which
the two--body fitting procedure is replaced by a series of cuts on kinematic 
distributions~\cite{brent}. This alternative analysis also yields a 
limit on $\tau_{\,\overline{p}}/B(\overline{p} \rightarrow e^-\pi^0)$ similar
to our quoted result.

\subsection{Search for 
$\overline{p} \rightarrow e^-\eta,\;e^-K^0_S,\;e^-\omega$, and 
$e^-\gamma\gamma$ \protect\label{eother}}

We begin by describing the search for $\overline{p} \rightarrow e^-\eta$. The
selection requirements employed for this decay mode are also used for the
search for the $e^-K^0_S,\;e^-\omega$, and $e^-\gamma\gamma$ final
states. The results for these latter decay modes are presented in
Sec.~\ref{ek0s}.

\subsubsection{$\overline{p} \rightarrow e^-\eta$ \protect\label{eeta}}

\begin{figure}
\vspace{-2.2cm}
\epsfxsize6.in
\centerline{\epsffile{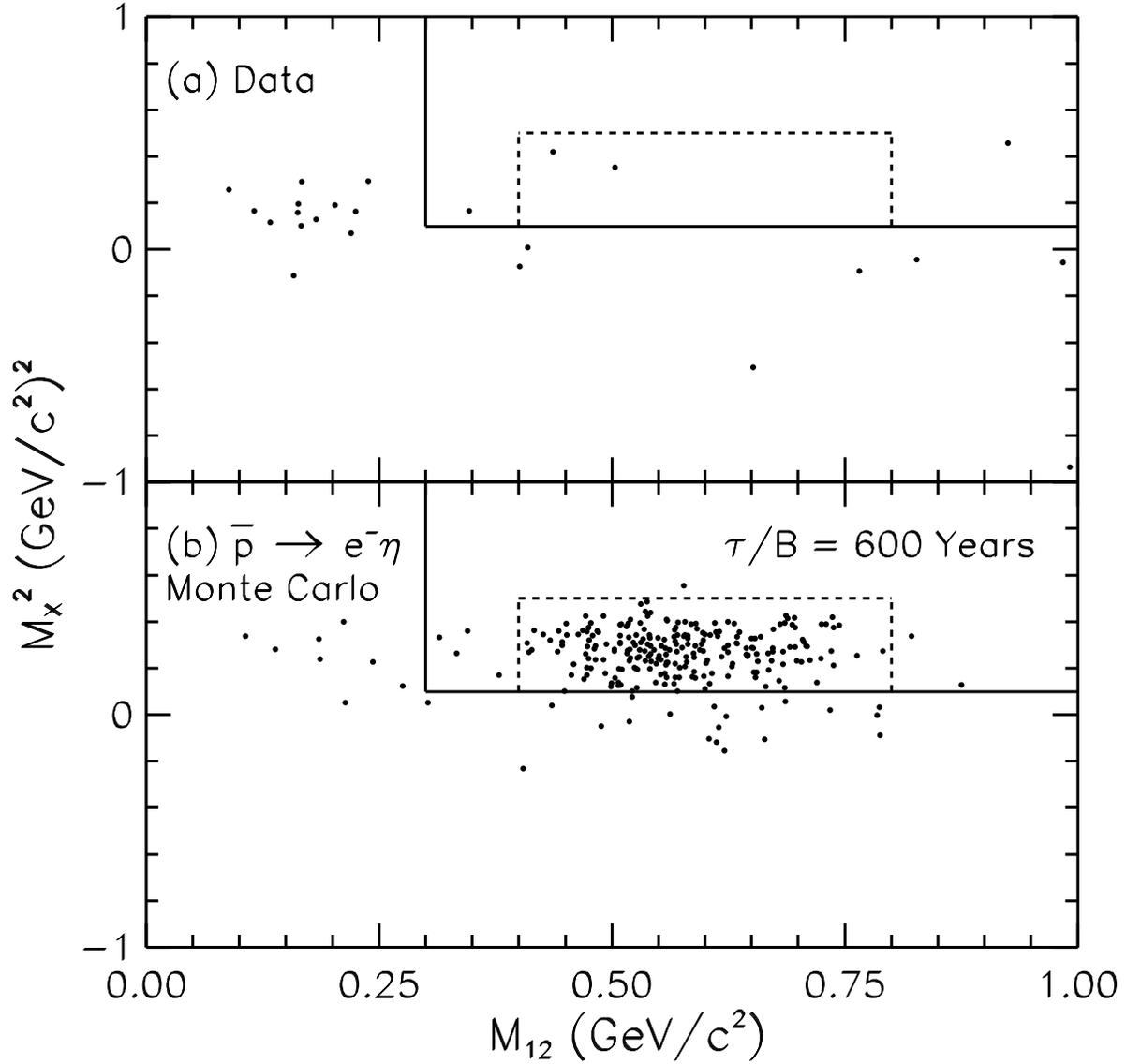}}
\vspace{-2.2cm}
\caption{Observed event populations in the (a) missing--mass--squared 
versus cluster--cluster mass plane for candidate events, and (b) simulated
$\overline{p} \rightarrow e^-\eta$ decays that pass the prior selection
requirements described in the text. The simulated data correspond to a lifetime
of 600~years.The dashed lines delineate the region accepted by the analysis
cuts for $\overline{p} \rightarrow e^-\eta$ decays and the solid lines
delineate the region accepted by the analysis cuts for
other modes, as described in the text.}
\label{eta_fig}
\end{figure}

The $\eta$ decays into neutral modes ($\eta \rightarrow 2\gamma,\; 3\pi^0,
\;\pi^02\gamma$) with a branching fraction of $\sim 0.7$. The decay
$\overline{p} \rightarrow e^-\eta$, with the $\eta$ subsequently decaying into
these neutral modes, would give rise to an electron plus two--or--more photons
in the final state. To select events consistent with the decay $\overline{p}
\rightarrow e^-\eta$ we begin with the \mbox{$\overline{p} \rightarrow e^{-}
\gamma$} selection requirements (1)--(9) described in Sec.~\ref{egamma}, which
yield the sample of $2.2 \times 10^4$ events in which there is a candidate
electron together with 1 or 2 additional clusters in the calorimeter. We then
require that:
\begin{description}
\item{(i)} There are two clusters in addition to the electron 
candidate ($1.2 \times 10^4$ events).
\item{(ii)} The lowest energy neutral cluster has an energy 
exceeding 1~GeV (1222 events), and the other two clusters 
both have energies in the range $2 < E < 6$~GeV (394 events).
\item{(iii)} The electron candidate is coplanar
($\Delta \phi > 2.8$~radians) with the beamline and the neutral
cluster pair (46 events). 
The coplanarity requirement is based upon a GEANT Monte Carlo study 
of simulated $\overline{p} \rightarrow e^-\eta$ decays within the 
APEX decay tank, and is less stringent 
than the corresponding requirements used for the 
$\overline{p} \rightarrow e^-\gamma$ and 
$\overline{p} \rightarrow e^-\pi^0$ 
searches described previously. 
\item{(iv)} The total energy measured in the calorimeter is 
in the range $7 < E_{tot} < 13$~GeV (30 events).
\item{(v)} The transverse--momentum--balance variable $p_T^{bal}$ 
is less than 0.3, where 
$p_T^{bal} \equiv | p_t(e) - p_t(\gamma\gamma) | / 
( p_t(e) + p_t(\gamma\gamma))$ (24 events).
\item{(vi)} The event lies within the dashed--region of the 
($M_X^2$, $M_{12}$)--plane shown in Fig.~\ref{eta_fig} (2 events), where 
the missing--mass--squared ($M_X^2$) is computed using the incoming 
$\overline{p}$ four--vector and the outgoing electron candidate 
four--vector, and $M_{12}$ is the measured mass of the cluster--pair. 
The region accepted contains 86\% of the simulated 
$\overline{p} \rightarrow e^-\eta$ decays that survive the previous 
selection requirements.
\end{description}

The distribution of the observed events in the ($M_X^2$, $M_{12}$)--plane is
shown in Fig.~\ref{eta_fig}a for the 24 events that survive the selection
criteria up to (but not including) the ($M_X^2$, $M_{12}$)--requirement. The
event population suggests that the 2 events that survive the ($M_X^2$,
$M_{12}$)--requirement are associated with the high--mass tail of the
background distribution. However, we cannot distinguish these events from
genuine $\overline{p} \rightarrow e^-\eta$ decays, and hence our limit on this
decay mode is based on the observation of 2 events, yielding $\mu_{max} = 5.7$
at 90\% C.L. The efficiency for $\overline{p} \rightarrow e^-\eta$ decays to
pass the trigger and offline selection requirements has been evaluated using
the GEANT Monte Carlo to generate $6 \times 10^4$ simulated decays uniformly
along the beamline within the decay tank. We find $\epsilon = (4.4 \pm 1.1)
\times 10^{-5}$, where the uncertainty on $\epsilon$ arises from the systematic
uncertainties on the trigger calibration, calorimeter energy scale, track
reconstruction efficiency, and Monte Carlo statistics. Inserting the calculated
$\epsilon$ and $\sigma_\epsilon$ into Eqs.~3 and 5 we obtain the result 
$\tau_{\,\overline{p}}/B(\overline{p} \rightarrow e^-\eta) \; > \;$ $2 \times
10^4$~yr (90\% C.L.).

\subsubsection{$\overline{p} \rightarrow 
e^-K^0_S,\;e^-\omega$, and 
$e^-\gamma\gamma$ \protect\label{ek0s}}

The selection requirements defined for the $\overline{p} \rightarrow e^-\eta$
search will also select candidates for other two--body $\overline{p}$ decay 
modes with an electron in the final state, provided the neutral particle
accompanying the electron (i) is sufficiently massive to produce events that
satisfy an ($M_X^2$, $M_{12}$)--requirement, and (ii) decays at least some of 
the time into daughter particles that are observed as two neutral clusters. In
particular, the $\overline{p} \rightarrow e^-\eta$ search results enable us to
place limits on the decay modes $\overline{p} \rightarrow
e^-K^0_S,\;e^-\omega$, and $e^-\gamma\gamma$. For these modes we
require candidate events fall in the region defined by $0.1 < M_X^2 <
1.0$~GeV$^2$/c$^4$ and $0.3 < M_{12} < 1.0$~GeV/c$^2$ (Fig.~\ref{eta_fig}). 
The corresponding limits will be based upon the observation of 4 events,
yielding $\mu_{max} = 8.9$ at 90\% C.L. Note that, of the simulated signal
events that pass all the prior selection requirements, the final ($M_X^2$,
$M_{12}$)--requirement accepts 45\% of the simulated $e^-K^0_S$ decays, 70\%
of the $e^-\omega$ decays, and 57\% of the
$\overline{p} \rightarrow e^-\gamma\gamma$ decays. The efficiencies for these
decays to pass the trigger and offline selection requirements have been
evaluated using the GEANT Monte Carlo program, and are given by $\epsilon =
(3.0 \pm 0.9) \times 10^{-6}$ for the $e^-K^0_S$ mode, $\epsilon = (6.8 \pm
1.9) \times 10^{-7}$ for the $e^-\omega$ mode, and 
$\epsilon = (4.9 \pm 1.3) \times
10^{-5}$ for the $e^-\gamma\gamma$ mode. Inserting the calculated values for
$\epsilon$ and $\sigma_\epsilon$ into Eqs.~3 and 5, at 90\% C.L. we obtain the
results $\tau_{\,\overline{p}}/B(\overline{p} \rightarrow e^-X) \; > \;$
900~yr for the $e^-K^0_S$ mode, 200~yr for the $e^-\omega$ mode, and 
$2 \times 10^4$~yr for the $e^-\gamma\gamma$ mode.

\subsection{Search for $\overline{p} \rightarrow e^-K^0_L$ 
\protect\label{ek0l}}

A $K^0_L$ with an energy of a few GeV will travel of order 100~m before
decaying. Hence, $\overline{p} \rightarrow e^-K^0_L$ decays occurring within
the APEX fiducial volume would be expected to produce an electron accompanied
by one neutral cluster in the calorimeter. In our search for $\overline{p}
\rightarrow e^-K^0_L$ decays we begin with the \mbox{$\overline{p} \rightarrow
e^{-} \gamma$} selection requirements (1)--(9) described in Sec.~\ref{egamma},
which yield the sample of $2.2 \times 10^4$ events in which there is a
candidate electron together with 1 or 2 additional clusters in the calorimeter.
We then require that:
\begin{description}
\item{(i)} There is only one cluster in addition to the electron 
candidate (9300 events).
\item{(ii)} The electron candidate is coplanar
($\Delta \phi > 3$~radians) with the beamline and the neutral
cluster (300 events). 
\item{(iii)} The total energy measured in the calorimeter is 
in the range $7 < E_{tot} < 13$~GeV (157 events).
\item{(iv)} The electron cluster and the neutral cluster both 
have energies in the range $2 < E < 6$~GeV (11 events).
\item{(v)} The missing--mass--squared lies in the range 
$0.1 < M_X^2 < 0.4$~GeV$^2$/c$^4$, where 
$M_X^2$ is computed using the incoming 
$\overline{p}$ four--vector and the outgoing electron candidate 
four--vector (3 events).
\end{description}

\begin{figure}
\vspace{-2.2cm}
\epsfxsize6.in
\centerline{\epsffile{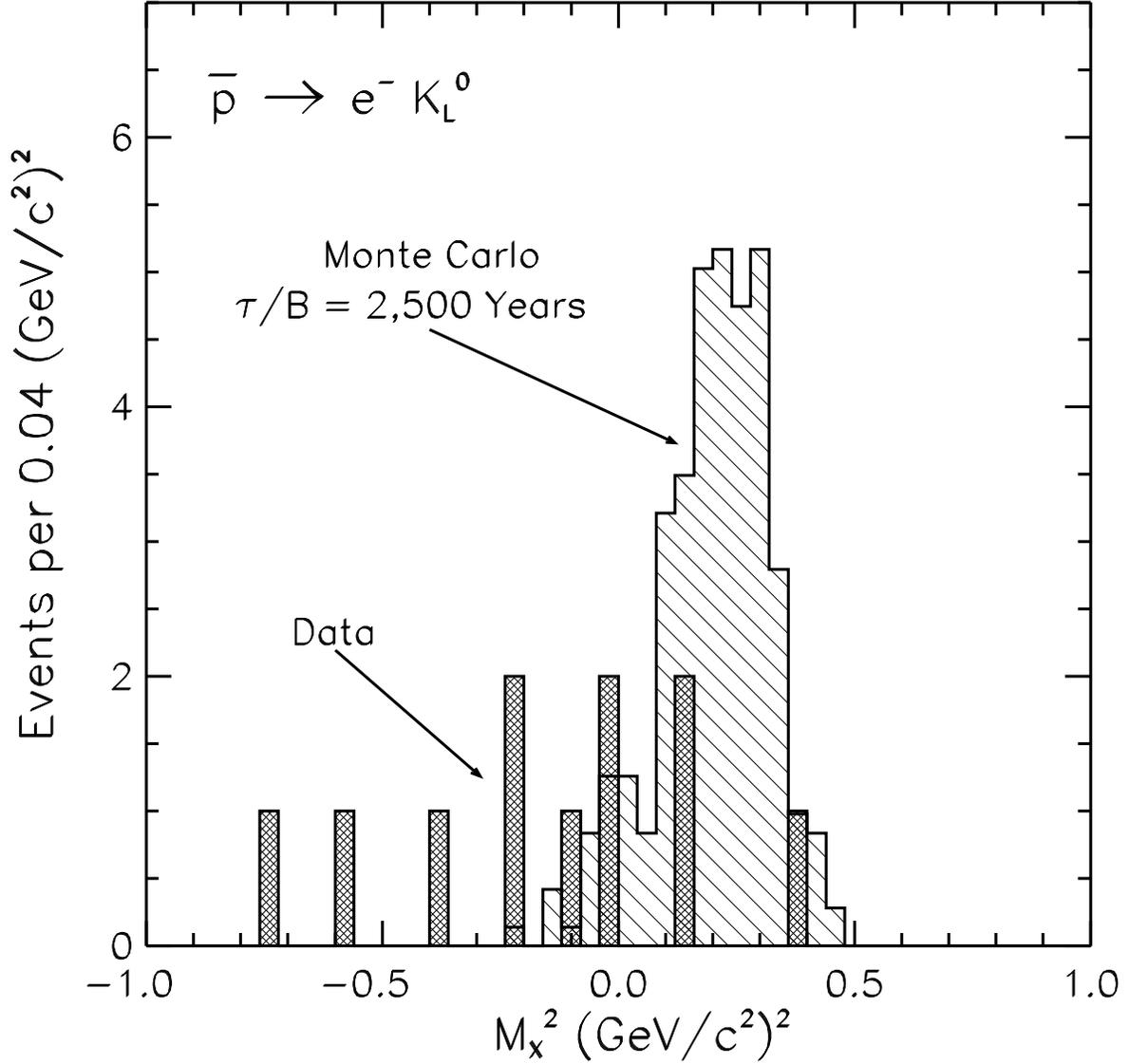}}
\vspace{-2.2cm}
\caption{Distribution of missing--mass--squared for the 11 
$\overline{p} \rightarrow e^-K^0_L$ candidate events 
that survive all of the requirements described 
in the text except for the missing--mass--squared requirement. 
The observed distributions are compared
with the GEANT Monte Carlo predictions for 
$\overline{p} \rightarrow e^-K^0_L$ decays 
normalized to correspond 
to a lifetime of 2,500~years.}
\label{kl_fig}
\end{figure}

The $M_X^2$--distribution for the observed events is shown in
Fig.~\ref{kl_fig} for the 11 events that survive the selection criteria up to
(but not including) the $M_X^2$ requirement. The event population suggests
that the 3 events that survive the $M_X^2$ requirement are associated with
the high--mass tail of the background distribution. However, we cannot
distinguish these events from genuine $\overline{p} \rightarrow e^-K^0_L$
decays, and hence our limit on this decay mode is based on the observation of
3 events, yielding $\mu_{max} = 7.3$ at 90\% C.L. The efficiency for
$\overline{p} \rightarrow e^-K^0_L$ decays to pass the trigger and offline
selection requirements has been evaluated using the GEANT Monte Carlo to
generate $10^5$ simulated decays uniformly along the beamline within the decay
tank. We find $\epsilon = (2.2 \pm 0.5) \times 10^{-5}$, where the uncertainty
on $\epsilon$ arises from the systematic uncertainties on the trigger
calibration, calorimeter energy scale, track reconstruction efficiency, and
Monte Carlo statistics. Inserting the calculated $\epsilon$ and
$\sigma_\epsilon$ into Eqs.~3 and 5 we obtain the result 
$\tau_{\,\overline{p}}/B(\overline{p} \rightarrow e^-K^0_L) \; > \;$ $9 \times
10^3$~yr (90\% C.L.).

\section{Conclusions \protect\label{conclusions}}

A search has been made for 11 two--body and 2 three-body decay modes of the
antiproton in which there is an electron or muon in the final state. No
statistically significant signal indicating $\overline{p}$ decay has been
observed. Our limits on $\tau/B$ are summarized in Tab.~\ref{results_tab},
and are significantly more stringent than the limits currently quoted by the
Particle Data Group~\cite{protondecay}. In particular, our results place the
first explicit experimental limits on the muonic decay modes of the
antiproton, and the first limits on the decay modes $e^-\gamma\gamma$ and
$e^-\omega$. Our most stringent limit is on
the decay mode $\overline{p} \rightarrow e^-\gamma$ for which we find $\tau/B
> 7 \times 10^5$~yr (90\% C.L.), which is an improvement by a factor of 400
over the prior T861 result. Noting that our limits for the simplest two--body
decay modes are in the range $10^4$ to $10^5$~years, Eq.~\ref{mxlimit}
suggests that if there is a CPT violating interaction that gives rise to these
decay modes, and that is described by a dimension--6 operator, the associated 
mass scale $m_X$ is greater than O($10^9$)~GeV/c$^2$. This is well beyond the 
reach of direct searches at high energy colliders.

\section{Acknowledgments}

The APEX experiment was performed at the Fermi National Accelerator 
Laboratory, which is operated by Universities Research Association, under
contract DE-AC02-76CH03000 with the U.S. Department of Energy.

%
%
\clearpage

\clearpage

\begin{table}
\centering
\caption{Summary of results: 90\% C.L. limits on $\tau / B$ for 13 antiproton 
decay modes.}
\vspace{0.2cm}
\begin{tabular}{cccc}
Decay Mode &$\tau/B$ Limit & Decay Mode &$\tau/B$ Limit\\
           & (years) &  & (years) \\
\hline
$\mu^-\gamma$       & $5 \times 10^4$ & $e^-\gamma$      & $7 \times 10^5$\\
$\mu^-\pi^0$        & $5 \times 10^4$ & $e^-\pi^0$       & $4 \times 10^5$\\
$\mu^-\eta$         & $8 \times 10^3$ & $e^-\eta$        & $2 \times 10^4$\\
$\mu^-\gamma\gamma$ & $2 \times 10^4$ & $e^-\gamma\gamma$ & $2 \times 10^4$\\
$\mu^-K^0_L$        & $7 \times 10^3$ & $e^-K^0_L$       & $9 \times 10^3$\\
$\mu^-K^0_S$        & $4 \times 10^3$ & $e^-K^0_S$       & $9 \times 10^2$\\
                    &                 & $e^-\omega$      & $2 \times 10^2$\\
\end{tabular}
\protect\label{results_tab}
\end{table}

%
%

\clearpage

\end{document}